# TENSOR PRODUCTS OF MODULES FOR A VERTEX OPERATOR ALGEBRA AND VERTEX TENSOR CATEGORIES

YI-ZHI HUANG AND JAMES LEPOWSKY

## 1. INTRODUCTION

In this paper, we present a theory of tensor products of classes of modules for a vertex operator algebra. We focus on motivating and explaining new structures and results in this theory, rather than on proofs, which are being presented in a series of papers beginning with [HL4] and [HL5]. An announcement has also appeared [HL1]. The theory is based on both the formal-calculus approach to vertex operator algebra theory developed in [FLM2] and [FHL] and the precise geometric interpretation of the notion of vertex operator algebra established in [H1].

Recently, mathematicians have been more and more attracted to conformal field theory, a physical theory which plays an important role in both condensed matter physics and string theory. Much of the research on conformal field theory has been centered on the conformal field theories determined by holomorphic fields of weight 1—the theories associated with certain highest-weight representations of affine Kac-Moody algebras. Many important structures and concepts which have arisen in recent years are related to this special class of conformal field theories—Wess-Zumino-Novikov-Witten (WZNW) models, the Knizhnik-Zamolodchikov equations and associated monodromy, quantum groups, braided tensor categories, the Jones polynomial and generalizations, three-manifold invariants, Chern-Simons theory, the Verlinde formula, etc. (see for instance [Wi1], [KZ], [J], [K1], [K2], [Ve], [TK], [MS], [Wi2], [TUY], [Dr1], [Dr2], [RT], [SV], [Va], [KL1]–[KL5], [Fi], [Fa]). But there are also other important mathematical structures that can be studied as conformal-field-theoretic structures, in particular, highest weight representations of the Virasoro algebra, $\mathcal{W}$-algebras and their representations, and most particularly for us, the moonshine module $V^\natural$ for the Fischer-Griess Monster sporadic finite simple group [Gr] constructed in [FLM1] and [FLM2]. For the conformal field theories associated with these structures, there are no nonzero holomorphic fields of weight 1, and correspondingly, the special methods for studying those conformal field theories associated with affine Lie algebras do not apply. We must have a broader viewpoint than that of this special class of conformal field theories.

The theory of vertex operator algebras, developed in [B1], [FLM2] and [FHL] (the research monograph [FLM2] also includes a detailed exposition of the theory, with examples), provides such a framework. Vertex operator algebras are "complex analogues" of both Lie algebras and commutative associative algebras. All the structures mentioned in the preceding paragraph can be and have been illuminated by the representation theory of vertex operator algebras. One important class of vertex operator algebras, generated by vectors of weight 1, is constructed from certain highest weight modules for affine Lie algebras, and certain other highest weight modules can be given the structure of modules for these vertex operator algebras (see for example [FZ], [DL]). The theories of highest weight representations of the Virasoro algebra and of representations of $\mathcal{W}$-algebras can also be studied in terms of the representation theory of the corresponding vertex operator algebras.

The moonshine module $V^\natural$ [FLM1] is in fact an early example of vertex operator algebra, except that a full vertex operator algebra structure on $V^\natural$ was stated to exist only later, by Borcherds in the announcement [B1], in which the notion of "vertex algebra" was introduced. In these terms, the structure detailed in the announcement [FLM1] was the Monster-invariant generating weight-two substructure, equipped correspondingly with a "cross-bracket" operation (rather than the Lie bracket operation), of the vertex operator algebra structure on $V^\natural$. The proof of the properties of the construction of $V^\natural$ and of the action of the Monster on it [FLM1] was given in [FLM2], along with the construction of a Monster-invariant vertex operator algebra structure on it (stated to exist in [B1]), and an axiomatic study of the concept of vertex operator algebra was presented in the monograph [FHL]. Meanwhile, Belavin, Polyakov and Zamolodchikov [BPZ] and other physicists also introduced the basic features (on a physical level of rigor) of what mathematicians came to understand as vertex operator algebras.

The most natural viewpoint for us is the viewpoint based on what we call the "Jacobi identity" (see [FLM2], [FHL] and the exposition [L1]) for vertex operator algebras. A "complex analogue" of the Jacobi identity for Lie algebras, this identity is the main tool in the formal-calculus approach to the theory of vertex operators algebras developed in [FLM2] and [FHL] and used in the present work. It can be deduced from the physically-formulated axioms in [BPZ] or from the axioms for vertex algebras in [B1]. The notion of vertex operator algebra of [FLM2] and [FHL] is in fact a variant of the notion of vertex algebra of [B1], equipped with the Jacobi identity as the main axiom and with certain grading restrictions assumed. It is this variant that we need. The Jacobi identity expresses an infinite family of generalized commutators (products like the $Z$-algebra products of [LW] or the cross-brackets of [FLM1] mentioned above) in closed form via a generating function based on formal delta-functions. In particular, the Jacobi identity exhibits the Lie-algebra-like properties of a vertex operator algebra rather than its associative-algebra-like properties, which are implicit in the operator-product-expansion formalism in the physics literature, including [BPZ], and in the axioms for vertex algebras in [B1].

The structure $V^\natural$ is a vertex operator algebra which is simple (irreducible as a module for itself) such that

(1) the rank (central charge) of $V^\natural$ is 24,
(2) there are no nonzero vectors of weight 1 in $V^\natural$,
(3) $V^\natural$ is its only irreducible module.

All these properties were established in [FLM2], except for the third, which was conjectured in [FLM2] and proved in [Do]. It was also conjectured in [FLM2] (and it remains a conjecture) that that these three properties characterize the vertex operator algebra $V^\natural$. We know that on the one hand,

(1) ([FLM1], [FLM2]) the automorphism group of $V^\natural$ is the Monster,

while on the other hand,

(2) [Do] every $V^\natural$-module is a direct sum of copies of $V^\natural$ itself (cf. (3) above), and in particular, the fusion algebra of $V^\natural$ is the trivial one-dimensional associative algebra and all the fusing and braiding matrices are trivial for the associated holomorphic conformal field theory.

We have a curiously unsatisfactory, or rather, unstable, situation—the same structure $V^\natural$ is the simplest possible from one viewpoint (that of monodromy of correlation functions) and the richest possible from another (that of symmetry); the Monster is arguably the most exceptional symmetry group that nature allows (cf. [Go]). We need a general theory to unify these two phenomena. (This problem was emphasized in [L2].) The theory of tensor products of modules for a vertex operator algebra, discussed in this paper, is being developed with this in mind.

Along with the algebraic theory of vertex operator algebras, our theory also has a geometric foundation: the techniques and results entering into the geometric interpretation of the notion of vertex operator algebra, in terms of a certain moduli space of multipunctured Riemann spheres with local coordinates vanishing at the punctures and the appropriate sewing operation on this moduli space, carried out in [H1]. The idea is to exploit fully the conformal structure implicit in the theory of vertex operator algebras, which can be thought of as the main ingredient of conformal field theory. A more conceptual reformulation of this geometric interpretation has been accomplished (see [HL2], [HL3]) with the help of the notion of operad, which was originally introduced ([S1], [S2], [M]) in a topological context in connection with the homotopy-theoretic characterization of loop spaces.

In the representation theory of Lie algebras, we have the classical notion of tensor product of modules, providing the conceptual foundation of the Clebsch-Gordan coefficients. The tensor product operation in the category of modules for a Lie algebra gives a classical example of a symmetric tensor category. Module categories for not-necessarily-cocommutative quantum groups (Hopf algebras) are sources of more general braided monoidal categories, which give rise to braid group representations and knot and link invariants; see in particular [J], [K1], [K2], [Dr1], [Dr2], [RT]. For vertex operator algebras, we also have the notions of modules, intertwining operators ("chiral vertex operators") among triples of modules and fusion rules analogous to those for Lie algebras (see [BPZ], [FS], [KZ], [TK], [Ve] and [MS]). The precise notion of intertwining operator that we need for the general theory, and that suggests the clearest analogy between the representation theory of vertex operator algebras and the representation theory of Lie algebras, is the one defined in [FHL], based on the Jacobi identity axiom for vertex operator algebras, rather than the more-commonly-used notion of chiral vertex operator based on certain Lie algebra coinvariants. One can ask whether there is a conceptual notion of tensor product of modules for a vertex operator algebra, providing a foundation for the notion of fusion rules, by analogy with the case of Lie algebras. The Jacobi identity axiom suggests a kind of "complex analogue" of a Hopf algebra diagonal map, but it turns out that a considerable amount of work is needed to make this idea precise.

Motivated partly by this point of view and partly by the announcement [KL1] of Kazhdan and Lusztig, conceptually constructing a tensor product operation and establishing its basic categorical properties for certain categories of modules for affine Lie algebras, we have introduced a general, conceptual notion of tensor products of modules for a suitable vertex operator algebra ([HL1], [HL4], [HL5]), and in a continuing series of papers, we are establishing its basic categorical properties. In place of the braided monoidal categories that arise from the Kazhdan-Lusztig construction ([KL1]–[KL5]) and its extension to the WZNW-model case by Finkelberg [Fi], the result is instead "vertex tensor categories," which are categories equipped with a suitably commutative tensor product operation controlled by the sewing of elements of the moduli space mentioned above rather than by the traditional commutative diagrams of coherence theory for monoidal categories. What we have is in some sense a conceptual "complexification" of the notion of symmetric monoidal category, in that the circle operad (cf. [HL2], [HL3]), which may be thought of as giving rise to the notion of symmetric monoidal category, is being replaced by a (partial) operad based analogously on the Riemann sphere. Moreover, a systematic specialization process yields an ordinary braided monoidal category from the vertex tensor category, for a suitable vertex operator algebra. This category is the usual braided monoidal category associated with the monodromy of correlation functions (essentially products of intertwining operators) and the braiding and fusing matrices, giving the connection with the representation theory of quantum groups, as in [KL1]–[KL5], [Fi], and with knot and link invariants. That is, the familiar and fundamental topological information generated by holomorphic conformal field theory at genus zero now becomes a specialization of a theory starting from an underlying conformal geometry.

Our approach is based on the general concepts of vertex operator algebra theory rather than the methods of [KL1]–[KL5] and [Fi], which use special properties of affine Lie algebras, and as we have mentioned, we need to use the notion of intertwining operator based on the Jacobi identity rather than the notion based on Lie algebra coinvariants; the two notions are indeed equivalent for the WZNW model and related models. Using this identity, we introduce a canonical notion of tensor product of modules for a suitable vertex operator algebra, defined in terms of an appropriate universal property and depending on a given element of the moduli space mentioned above. One aspect of this construction is that, as was also the case in [KL1]–[KL5], the underlying vector space of the tensor product module is not at all the tensor product vector space of the given modules. However, the theory does in fact provide an analogue of the concrete elements of ordinary classical tensor product modules (the usual linear combinations of "tensors" of elements of the given modules), namely, the space of elements of the dual space of the tensor product vector space of two modules satisfying a certain list of conditions, the most important of which is what we call the "compatibility condition," which is motivated by the Jacobi identity and which allows the abstract machinery to work. In fact, one of the main theorems in [HL1], [HL4] and [HL5] is that this space of vectors is in fact a module in a certain generalized sense; the proof of this result requires formal calculations based on the Jacobi identity. The desired tensor product module is then the contragredient module of this generalized module, in the case in which this generalized module is a module. This theorem enables us to establish the conceptual vertex-tensor-categorical properties of the tensor product operation, by analogy with the way in which one's ability to write down concrete tensor product vectors in a classical tensor product module enables one to establish the classical tensor-categorical properties (such as the associativity or commutativity properties). The machinery of both formal calculus and of the geometric interpretation of the notion of vertex operator algebra enter heavily into these considerations.

It should be emphasized that our theory applies (at present) to an arbitrary "rational" vertex operator algebra satisfying a certain convergence condition, including the vertex operator algebras associated with the WZNW models (see [KZ], [TK], [FZ], [DL] for the necessary properties) and with minimal models (see [BPZ], [H4], [FZ], [Wa]), and especially including the moonshine module (see [FLM2], [Do]). Even though the moonshine module exhibits no monodromy, it is expected to possess rich vertex-tensor-categorical structure coming from the conformal geometry. In particular, combined with some other structures (e.g., "vertex categories," which are analogues of categories, and "modular vertex tensor categories," which are higher-genus generalizations of vertex tensor categories) the present theory is expected to provide a resolution of the philosophical paradox described above and to shed light on such phenomena as "monstrous moonshine" (see [CN], [FLM1], [B1], [MN], [FLM2], [B2]). In the theory of tensor categories, we have the Tannaka-Krein reconstruction theorem (see for example [JS2]). For vertex tensor categories, we expect that a vertex tensor category together with certain additional structures determines uniquely (up to isomorphism) a vertex operator algebra such that the vertex tensor category constructed from a suitable category of modules for it is equivalent (in the sense of vertex tensor categories) to the original vertex tensor category. This is supported by the fact that the vertex tensor categories contructed from the moonshine module $V^\natural$ and from the trivial vertex operator algebra are not equivalent, even though the corresponding braided tensor categories are equivalent (and are both trivial). We also expect that this theory will give a better understanding of the conformal field theories constructed from representations of $\mathcal{W}$-algebras. In addition, many of the notions, constructions and techniques will also apply to vertex operator algebras which are not rational and module categories which are quite general—not necessarily semisimple.

In the present paper we give a brief description of our theory. The complete theory will be given in a series of papers beginning with [HL4] and [HL5]. For the reader's convenience, we review basic concepts in the representation theory of vertex operator algebras in Section 2. For a certain element $P(z)$ of the moduli space of spheres with three punctures and local coordinates, we describe the theory of $P(z)$-tensor products of two modules in Section 3. In Section 4, we introduce the notion of vertex tensor category, and we announce that the category of modules for a rational vertex operator algebra for which products of intertwining operators are convergent for variables in certain regions has the structure of a vertex tensor category. We also announce that any vertex tensor category gives rise to a braided tensor category.

Results and concepts in this paper were announced in talks presented by both authors at the June, 1992 AMS-IMS-SIAM Joint Summer Research Conference on Conformal Field Theory, Topological Field Theory and Quantum Groups at Mount Holyoke College.

**Acknowledgments.** We would like to thank D. Kazhdan, G. Lusztig and M. Finkelberg for interesting discussions, especially concerning the comparison between their approach and ours in the special case of affine Lie algebras. We are also grateful to I. M. Gelfand for initially directing our attention to the preprint of the paper [KL1] in his seminar at Rutgers University and to O. Mathieu for illuminating comments on that preprint. We thank R. Borcherds for informing us that some years ago, he also began considering a notion of tensor product of modules for a vertex algebra. During the course of this work, Y.-Z. H. has been supported in part by NSF grants DMS-8610730 (through the Institute for Advanced Study), DMS-9104519 and DMS-9301020 and J. L. by NSF grants DMS-8603151 and DMS-9111945 and the Rutgers University Faculty Academic Study Program. J. L. also thanks the Institute for Advanced Study for its hospitality.

## 2. REVIEW OF BASIC CONCEPTS

In this section, we review some basic definitions and concepts in the representation theory of vertex operator algebras. Everything in this section can be found in [FLM2] and [FHL], except for the following points: Here we use very slightly more general versions of the definitions of module and of intertwining operator than used in [FHL], allowing complex rather than rational gradings for modules. We also consider the notions of "opposite vertex operator" and of "generalized module."

In this paper, all the variables $x, x_0, \dots$ are independent commuting formal variables, and all expressions involving these variables are to be understood as formal Laurent series or, when explicitly so designated, as formal rational functions. (Later, we shall also use the symbol $z$, which will denote a complex number, *not* a formal variable.) We shall use the formal $\delta$-function $\delta(x) = \sum_{n \in \mathbb{Z}} x^n$ and the convention that negative powers of a binomial are to be expanded in nonnegative powers of the second summand, in expressions such as $\delta\left(\frac{x_1 - x_2}{x_0}\right)$, for example. See [FLM2] and [FHL] for extensive discussions of the calculus of formal $\delta$-functions, and see [HL4] for additional details concerning the subject of this paper.

We now quote the definition and basic "duality" properties of vertex operator algebras from [FLM2] or [FHL]:

**Definition 2.1.** A *vertex operator algebra* (over $\mathbb{C}$) is a $\mathbb{Z}$-graded vector space (graded by *weights*)

$$V = \coprod_{n \in \mathbb{Z}} V_{(n)}; \text{ for } v \in V_{(n)}, \ n = \text{wt } v; \tag{2.1}$$

such that

$$\dim V_{(n)} < \infty \text{ for } n \in \mathbb{Z}, \tag{2.2}$$

$$V_{(n)} = 0 \text{ for } n \text{ sufficiently small}, \tag{2.3}$$

equipped with a linear map $V \otimes V \to V[[x, x^{-1}]]$, or equivalently,

$$\begin{aligned} V &\to (\text{End } V)[[x, x^{-1}]] \\ v &\mapsto Y(v, x) = \sum_{n \in \mathbb{Z}} v_n x^{-n-1} \ (\text{where } v_n \in \text{End } V), \end{aligned} \tag{2.4}$$

$Y(v, x)$ denoting the *vertex operator associated with* $v$, and equipped also with two distinguished homogeneous vectors $\mathbf{1} \in V_{(0)}$ (the *vacuum*) and $\omega \in V_{(2)}$. The following conditions are assumed for $u, v \in V$: the *lower truncation condition* holds:

$$u_n v = 0 \text{ for } n \text{ sufficiently large} \tag{2.5}$$

(or equivalently, $Y(u, x)v \in V((x))$);

$$Y(\mathbf{1}, x) = 1 \ (1 \text{ on the right being the identity operator}); \tag{2.6}$$

the *creation property* holds:

$$Y(v, x)\mathbf{1} \in V[[x]] \text{ and } \lim_{x \to 0} Y(v, x)\mathbf{1} = v \tag{2.7}$$

(that is, $Y(v, x)\mathbf{1}$ involves only nonnegative integral powers of $x$ and the constant term is $v$); the *Jacobi identity* (the main axiom) holds:

$$x_0^{-1}\delta\left(\frac{x_1 - x_2}{x_0}\right)Y(u, x_1)Y(v, x_2) - x_0^{-1}\delta\left(\frac{x_2 - x_1}{-x_0}\right)Y(v, x_2)Y(u, x_1)$$
$$= x_2^{-1}\delta\left(\frac{x_1 - x_0}{x_2}\right)Y(Y(u, x_0)v, x_2) \tag{2.8}$$

(note that when each expression in (2.8) is applied to any element of $V$, the coefficient of each monomial in the formal variables is a finite sum; on the right-hand side, the notation $Y(\cdot, x_2)$ is understood to be extended in the obvious way to $V[[x_0, x_0^{-1}]]$); the *Virasoro algebra relations* hold:

$$[L(m), L(n)] = (m - n)L(m + n) + \frac{1}{12}(m^3 - m)\delta_{n+m,0}c \tag{2.9}$$

for $m, n \in \mathbb{Z}$, where

$$L(n) = \omega_{n+1} \text{ for } n \in \mathbb{Z}, \text{ i.e., } Y(\omega, x) = \sum_{n \in \mathbb{Z}} L(n)x^{-n-2} \tag{2.10}$$

and

$$c \in \mathbb{C}; \tag{2.11}$$

$$L(0)v = nv = (\text{wt } v)v \text{ for } n \in \mathbb{Z} \text{ and } v \in V_{(n)}; \tag{2.12}$$

$$\frac{d}{dx}Y(v, x) = Y(L(-1)v, x) \tag{2.13}$$

(the *$L(-1)$-derivative property*).

The vertex operator algebra just defined is denoted by $(V, Y, \mathbf{1}, \omega)$ (or simply by $V$). The complex number $c$ is called the *central charge* or *rank* of $V$. Homomorphisms of vertex operator algebras are defined in the obvious way.

Vertex operator algebras have important "rationality," "commutativity" and "associativity" properties, collectively called "duality" properties. These properties can in fact be used as axioms replacing the Jacobi identity in the definition of vertex operator algebra, as we now recall.

In the propositions below, $\mathbb{C}[x_1, x_2]_S$ is the ring of rational functions obtained by inverting (localizing with respect to) the products of (zero or more) elements of the set $S$ of nonzero homogeneous linear polynomials in $x_1$ and $x_2$. Also, $\iota_{12}$ (which might also be written as $\iota_{x_1 x_2}$) is the operation of expanding an element of $\mathbb{C}[x_1, x_2]_S$, that is, a polynomial in $x_1$ and $x_2$ divided by a product of homogeneous linear polynomials in $x_1$ and $x_2$, as a formal series containing at most finitely many negative powers of $x_2$ (using binomial expansions for negative powers of linear polynomials involving both $x_1$ and $x_2$); similarly for $\iota_{21}$ and so on. (The distinction between rational functions and formal Laurent series is crucial.)

For any $\mathbb{Z}$-graded, or more generally, $\mathbb{C}$-graded, vector space $W = \coprod W_{(n)}$, we use the notation

$$W' = \coprod W_{(n)}^* \tag{2.14}$$

for its graded dual. We denote the canonical pairing between $W'$ and $W$ by $\langle \cdot, \cdot \rangle$.

**Proposition 2.2. (a) (rationality of products)** *For $v$, $v_1$, $v_2 \in V$ and $v' \in V'$, the formal series $\langle v', Y(v_1, x_1) Y(v_2, x_2) v \rangle$, which involves only finitely many negative powers of $x_2$ and only finitely many positive powers of $x_1$, lies in the image of the map $\iota_{12}$:*

$$\langle v', Y(v_1, x_1) Y(v_2, x_2) v \rangle = \iota_{12} f(x_1, x_2), \tag{2.15}$$

*where the (uniquely determined) element $f \in \mathbb{C}[x_1, x_2]_S$ is of the form*

$$f(x_1, x_2) = \frac{g(x_1, x_2)}{x_1^r x_2^s (x_1 - x_2)^t} \tag{2.16}$$

*for some $g \in \mathbb{C}[x_1, x_2]$ and $r, s, t \in \mathbb{Z}$.*

**(b) (commutativity)** *We also have*

$$\langle v', Y(v_2, x_2) Y(v_1, x_1) v \rangle = \iota_{21} f(x_1, x_2). \tag{2.17}$$

**Proposition 2.3. (a) (rationality of iterates)** *For $v$, $v_1$, $v_2 \in V$ and $v' \in V'$, the formal series $\langle v', Y(Y(v_1, x_0) v_2, x_2) v \rangle$, which involves only finitely many negative powers of $x_0$ and only finitely many positive powers of $x_2$, lies in the image of the map $\iota_{20}$:*

$$\langle v', Y(Y(v_1, x_0) v_2, x_2) v \rangle = \iota_{20} h(x_0, x_2), \tag{2.18}$$

*where the (uniquely determined) element $h \in \mathbb{C}[x_0, x_2]_S$ is of the form*

$$h(x_0, x_2) = \frac{k(x_0, x_2)}{x_0^r x_2^s (x_0 + x_2)^t} \tag{2.19}$$

*for some $k \in \mathbb{C}[x_0, x_2]$ and $r, s, t \in \mathbb{Z}$.*

**(b)** *The formal series $\langle v', Y(v_1, x_0 + x_2) Y(v_2, x_2) v \rangle$, which involves only finitely many negative powers of $x_2$ and only finitely many positive powers of $x_0$, lies in the image of $\iota_{02}$, and in fact*

$$\langle v', Y(v_1, x_0 + x_2) Y(v_2, x_2) v \rangle = \iota_{02} h(x_0, x_2). \tag{2.20}$$

**Proposition 2.4 (associativity).** *We have the following equality of rational functions:*

$$\iota_{12}^{-1} \langle v', Y(v_1, x_1) Y(v_2, x_2) v \rangle = \left( \iota_{20}^{-1} \langle v', Y(Y(v_1, x_0) v_2, x_2) v \rangle \right) \Big|_{x_0 = x_1 - x_2}. \tag{2.21}$$

**Proposition 2.5.** *In the presence of the other axioms, the Jacobi identity follows from the rationality of products and iterates, and commutativity and associativity. In particular, in the definition of vertex operator algebra, the Jacobi identity may be replaced by these properties.*

We have the following notions of module and of intertwining operator for vertex operator algebras:

**Definition 2.6.** Given a vertex operator algebra $(V, Y, \mathbf{1}, \omega)$, a *module for $V$* (or *$V$-module* or *representation space*) is a $\mathbb{C}$-graded vector space (graded by *weights*)

$$W = \coprod_{n \in \mathbb{C}} W_{(n)}; \ \text{ for } \ w \in W_{(n)}, \ \ n = \text{wt } w; \tag{2.22}$$

such that

$$\dim \ W_{(n)} < \infty \ \text{ for } \ n \in \mathbb{C}, \tag{2.23}$$

$$W_{(n)} = 0 \ \text{ for } \ n \ \text{ whose real part is sufficiently small,} \tag{2.24}$$

equipped with a linear map $V \otimes W \to W[[x, x^{-1}]]$, or equivalently,

$$\begin{aligned} V &\to \ (\text{End } W)[[x, x^{-1}]] \\ v &\mapsto \ Y(v, x) = \sum_{n \in \mathbb{Z}} v_n x^{-n-1} \quad (\text{where } \ v_n \in \text{End } W) \end{aligned} \tag{2.25}$$

(note that the sum is over $\mathbb{Z}$, not $\mathbb{C}$), $Y(v, x)$ denoting the *vertex operator associated with $v$*, such that "all the defining properties of a vertex operator algebra that make sense hold." That is, for $u, v \in V$ and $w \in W$,

$$v_n w = 0 \ \text{ for } \ n \ \text{ sufficiently large} \tag{2.26}$$

(the lower truncation condition);

$$Y(\mathbf{1}, z) = 1; \tag{2.27}$$

$$\begin{aligned} x_0^{-1} \delta \left( \frac{x_1 - x_2}{x_0} \right) Y(u, x_1) Y(v, x_2) &- x_0^{-1} \delta \left( \frac{x_2 - x_1}{-x_0} \right) Y(v, x_2) Y(u, x_1) \\ &= x_2^{-1} \delta \left( \frac{x_1 - x_0}{x_2} \right) Y(Y(u, x_0)v, x_2) \end{aligned} \tag{2.28}$$

(the Jacobi identity for operators on $W$); note that on the right-hand side, $Y(u, x_0)$ is the operator associated with $V$; the Virasoro algebra relations hold on $W$ with scalar $c$ equal to the central charge of $V$:

$$[L(m), L(n)] = (m - n)L(m + n) + \frac{1}{12}(m^3 - m)\delta_{m+n, 0} c \tag{2.29}$$

for $m, n \in \mathbb{Z}$, where

$$L(n) = \omega_{n+1} \ \text{ for } \ n \in \mathbb{Z}, \ \text{ i.e., } \ Y(\omega, z) = \sum_{n \in \mathbb{Z}} L(n) x^{-n-2}; \tag{2.30}$$

$$L(0)w = nw = (\text{wt } w)w \ \text{ for } \ n \in \mathbb{C} \ \text{ and } \ w \in W_{(n)}; \tag{2.31}$$

$$\frac{d}{dx} Y(v, x) = Y(L(-1)v, x), \tag{2.32}$$

where $L(-1)$ is the operator on $V$.

We may denote the module just defined by $(W, Y)$ (or simply by $W$). If necessary, we shall use $Y_W$ or similar notation to indicate that the vertex operators concerned act on $W$. Homomorphisms (or maps) of $V$-modules are defined in the obvious way. For $V$-modules $W_1$ and $W_2$, we shall denote the space of module maps from $W_1$ to $W_2$ by $\text{Hom}_V(W_1, W_2)$.

For any vector space $W$ and any formal variable $x$, we use the notation

$$W\{x\} = \left\{ \sum_{n \in \mathbb{C}} a_n x^n | a_n \in W, \ n \in \mathbb{C} \right\}. \tag{2.33}$$

In particular, we shall allow complex powers of our commuting formal variables.

**Definition 2.7.** Let $V$ be a vertex operator algebra and let $(W_1, Y_1)$, $(W_2, Y_2)$ and $(W_3, Y_3)$ be three $V$-modules (not necessarily distinct, and possibly equal to $V$). An *intertwining operator of type* $\binom{W_3}{W_1 \, W_2}$ is a linear map $W_1 \otimes W_2 \to W_3\{x\}$, or equivalently,

$$\begin{aligned} W_1 &\to (\text{Hom}(W_2, W_3))\{x\} \\ w &\mapsto \mathcal{Y}(w, x) = \sum_{n \in \mathbb{C}} w_n x^{-n-1} \ \ (\text{where } \ w_n \in \text{Hom}(W_2, W_3)) \end{aligned} \tag{2.34}$$

such that "all the defining properties of a module action that make sense hold." That is, for $v \in V$, $w_{(1)} \in W_1$ and $w_{(2)} \in W_2$, we have the lower truncation condition

$$(w_{(1)})_n w_{(2)} = 0 \ \text{ for } \ n \ \text{ whose real part is sufficiently large;} \tag{2.35}$$

the following Jacobi identity holds for the operators $Y_1(v, \cdot)$, $Y_2(v, \cdot)$, $Y_3(v, \cdot)$ and $\mathcal{Y}(\cdot, x_2)$ acting on the element $w_{(2)}$:

$$\begin{aligned} x_0^{-1} \delta &\left( \frac{x_1 - x_2}{x_0} \right) Y_3(v, x_1) \mathcal{Y}(w_{(1)}, x_2) w_{(2)} \\ &- x_0^{-1} \delta \left( \frac{x_2 - x_1}{-x_0} \right) \mathcal{Y}(w_{(1)}, x_2) Y_2(v, x_1) w_{(2)} \\ &= x_2^{-1} \delta \left( \frac{x_1 - x_0}{x_2} \right) \mathcal{Y}(Y_1(v, x_0) w_{(1)}, x_2) w_{(2)} \end{aligned} \tag{2.36}$$

(note that the first term on the left-hand side is algebraically meaningful because of the condition (2.35), and the other terms are meaningful by the usual properties of modules; also note that this Jacobi identity involves integral powers of $x_0$ and $x_1$ and complex powers of $x_2$);

$$\frac{d}{dx}\mathcal{Y}(w_{(1)}, x) = \mathcal{Y}(L(-1)w_{(1)}, x),$$ (2.37)

where $L(-1)$ is the operator acting on $W_1$.

The intertwining operators of the same type $\binom{W_3}{W_1\,W_2}$ form a vector space, which we denote by $\mathcal{V}^{W_3}_{W_1 W_2}$. The dimension of this vector space is called the *fusion rule* for $W_1$, $W_2$ and $W_3$ and is denoted by $N^{W_3}_{W_1 W_2}$ $(\leq \infty)$.

There are also duality properties for modules and intertwining operators. See [FHL] and [DL] for details.

For any $V$-module $(W, Y)$, we define the *opposite vertex operator* associated to $v \in V$ by

$$Y^*(v, x) = Y(e^{xL(1)}(-x^{-2})^{L(0)}v, x^{-1}),$$ (2.38)

as in [HL4]. Then $Y^*$ satisfies the following *opposite Jacobi identity*:

$$x_0^{-1}\delta\left(\frac{x_1 - x_2}{x_0}\right) Y^*(v_2, x_2) Y^*(v_1, x_1)$$
$$-x_0^{-1}\delta\left(\frac{x_2 - x_1}{-x_0}\right) Y^*(v_1, x_1) Y^*(v_2, x_2)$$
$$= x_2^{-1}\delta\left(\frac{x_1 - x_0}{x_2}\right) Y^*(Y(v_1, x_0)v_2, x_2)$$ (2.39)

for $v_1, v_2 \in V$. The pair $(W, Y^*)$ should be thought of as a "right module" for $V$. We have $Y^{**} = Y$ (using formula (2.38) a second time to define $Y^{**}$).

Let $(W, Y)$, with

$$W = \coprod_{n \in \mathbb{C}} W_{(n)},$$ (2.40)

be a $V$-module, and consider its graded dual space $W'$ (recall (2.14)). We define the *contragredient vertex operators* (called "adjoint vertex operators" in [FHL]) $Y'(v, x)$ $(v \in V)$ by means of the linear map

$$V \rightarrow (\text{End } W')[[x, x^{-1}]]$$
$$v \mapsto Y'(v, x) = \sum_{n \in \mathbb{Z}} v'_n x^{-n-1} \quad (\text{where } v'_n \in \text{End } W'),$$ (2.41)

determined by the condition

$$\langle Y'(v, x)w', w\rangle = \langle w', Y^*(v, x)w\rangle$$ (2.42)

for $v \in V$, $w' \in W'$, $w \in W$.

We give the space $W'$ a $\mathbb{C}$-grading by setting

$$W'_{(n)} = W^*_{(n)} \ \text{ for } \ n \in \mathbb{C} \tag{2.43}$$

(cf. (2.14)). The following theorem defines the $V$-module $W'$ contragredient to $W$ (see [FHL], Theorem 5.2.1 and Proposition 5.3.1):

**Theorem 2.8.** *The pair* $(W', Y')$ *carries the structure of a $V$-module and* $(W'', Y'') = (W, Y)$.

Given a module map $\eta : W_1 \to W_2$, there is a unique module map $\eta' : W'_2 \to W'_1$, the *adjoint* map, such that

$$\langle \eta'(w'_{(2)}), w_{(1)} \rangle = \langle w'_{(2)}, \eta(w_{(1)}) \rangle \tag{2.44}$$

for $w_{(1)} \in W_1$ and $w_{(2)} \in W_2$. (Here the pairings $\langle \cdot, \cdot \rangle$ on the two sides refer to two different modules.) Note that

$$\eta'' = \eta. \tag{2.45}$$

In the construction of the tensor product module of two modules for a vertex operator algebra, we shall need the following generalization of the notion of module recalled above:

**Definition 2.9.** A *generalized $V$-module* is a $\mathbb{C}$-graded vector space $W$ equipped with a linear map of the form (2.25) satisfying all the axioms for a $V$-module except that the homogeneous subspaces need not be finite-dimensional and that they need not be zero even for $n$ whose real part is sufficiently small; that is, we omit (2.23) and (2.24) from the definition.

## 3. The notion of $P(z)$-tensor product and two constructions

The symbol $P(z)$ in the title of this section represents a geometric object associated with a fixed nonzero complex number $z$. Geometrically, to define a tensor product of modules for a vertex operator algebra, we need to specify an element of the moduli space $K$ of spheres with ordered punctures, the 0th puncture being negatively oriented and the others positively oriented, and with local coordinates vanishing at these punctures. More precisely, we need an element of the determinant line bundle over $K$ raised to the power $c$, where $c \in \mathbb{C}$ is the central charge of the vertex operator algebra. For detailed discussions of the moduli space $K$ and its crucial role in the geometric interpretation of the notion of vertex operator algebra, see [H1], [H2], [H3] [H4]. In this paper, we give the notion and two constructions of tensor product associated with the particular element $P(z)$ of $K$ containing $\mathbb{C} \cup \{\infty\}$ with ordered punctures $\infty$, $z$, 0 ($\infty$ being the (negatively oriented) 0th puncture and $z$ and 0 the (positively oriented) first and second punctures, respectively) and standard local coordinates $1/w$, $w - z$, $w$, vanishing at $\infty$, $z$, 0, respectively. This element $P(z)$ is the geometric object corresponding to vertex operators or intertwining operators in the geometric interpretation of vertex operators and intertwining operators. The tensor products associated with other elements of $K$ (or more precisely, elements of the determinant line bundle over $K$ raised to the power $c$) can be defined analogously and can be constructed using the results described here for the constructions of the $P(z)$-tensor product.

Before giving the definition of $P(z)$-tensor product of two modules for a vertex operator algebra, we first recall the version of the definition of tensor product of two modules for a Lie algebra which will be our main guide. In the theory of Lie algebras we have the following standard notion of intertwining map (of type $\binom{W_3}{W_1 W_2}$) among modules $W_1, W_2, W_3$ for a Lie algebra $V$, with corresponding actions $\pi_1, \pi_2, \pi_3$ of $V$: a linear map $I$ from the tensor product vector space $W_1 \otimes W_2$ to $W_3$, satisfying the identity

$$\pi_3(v)I(w_{(1)} \otimes w_{(2)}) = I(\pi_1(v)w_{(1)} \otimes w_{(2)}) + I(w_{(1)} \otimes \pi_2(v)w_{(2)}) \tag{3.1}$$

for $v \in V$, $w_{(1)} \in W_1$, $w_{(2)} \in W_2$. This "Jacobi identity for intertwining maps" agrees with the Jacobi identity for $V$ when all three modules are the adjoint module. Let us call a *product* of $W_1$ and $W_2$ a third module $W_3$ equipped with an intertwining map $I$ of type $\binom{W_3}{W_1 W_2}$; we denote this by $(W_3, I)$. Then a tensor product of $W_1$ and $W_2$ is a product $(W_1 \otimes W_2, \otimes)$ such that given any product $(W_3, I)$, there exists a unique module map $\eta$ from $W_1 \otimes W_2$ to $W_3$ such that

$$I = \eta \circ \otimes. \tag{3.2}$$

Thus any tensor product of two given modules has the following property: The intertwining maps from the tensor product vector space of the two modules to a third module correspond naturally to the module maps from the tensor product module to the third module. Moreover, this universal property characterizes the tensor product module up to unique isomorphism.

Note that this version of the definition of tensor product of two Lie algebra modules is based on knowledge of the intertwining maps relating three modules. The point is that in vertex operator algebra theory, we can formulate a notion of intertwining map (related to but distinct from the notion of intertwining operator), and using this we can define the appropriate notion of tensor product of modules for a vertex operator algebra. This tensor product will not have the vector space tensor product as its underlying vector space. We proceed to describe how this works.

For any $\mathbb{C}$-graded vector space $W = \coprod W_{(n)}$ such that $\dim W_{(n)} < \infty$ for each $n \in \mathbb{C}$, we use the notation

$$\overline{W} = \prod_{n \in \mathbb{C}} W_{(n)} = W'^*, \tag{3.3}$$

where as above $'$ denotes the graded dual space and as usual $^*$ denotes the dual space of a vector space.

Let $(V, Y, \mathbf{1}, \omega)$ be a vertex operator algebra and $(W, Y)$ a $V$-module. It follows from the axioms that for any $v \in V$ and $n \in \mathbb{Z}$, there is a well-defined natural action of $v_n$ on $\overline{W}$. Moreover, for fixed $v \in V$, any infinite linear combination of the $v_n$ of the form $\sum_{n<N} a_n v_n$ $(a_n \in \mathbb{C})$ acts on $\overline{W}$ in a well-defined way.

Fix $z \in \mathbb{C}^{\times}$ and let $(W_1, Y_1)$, $(W_2, Y_2)$ and $(W_3, Y_3)$ be $V$-modules. By a $P(z)$-$intertwining$ $map$ $of$ $type$ $\binom{W_3}{W_1 W_2}$ we mean a linear map $F : W_1 \otimes W_2 \to \overline{W}_3$ satisfying the condition

$$x_0^{-1} \delta \left( \frac{x_1 - z}{x_0} \right) Y_3(v, x_1) F(w_{(1)} \otimes w_{(2)}) =$$
$$= z^{-1} \delta \left( \frac{x_1 - x_0}{z} \right) F(Y_1(v, x_0) w_{(1)} \otimes w_{(2)})$$
$$+ x_0^{-1} \delta \left( \frac{z - x_1}{-x_0} \right) F(w_{(1)} \otimes Y_2(v, x_1) w_{(2)}) \tag{3.4}$$

for $v \in V$, $w_{(1)} \in W_1$, $w_{(2)} \in W_2$ (cf. the identity (3.1) and also the Jacobi identity (2.36)). The preceding paragraph shows that the left-hand side of (3.4) is well defined. We denote the vector space of $P(z)$-intertwining maps of type $\binom{W_3}{W_1 W_2}$ by $\mathcal{M}[P(z)]_{W_1 W_2}^{W_3}$. A $P(z)$-$product$ $of$ $W_1$ $and$ $W_2$ is a $V$-module $(W_3, Y_3)$ equipped with a $P(z)$-intertwining map $F$ of type $\binom{W_3}{W_1 W_2}$. We denote it by $(W_3, Y_3; F)$ (or simply by $(W_3, F)$). Let $(W_4, Y_4; G)$ be another $P(z)$-product of $W_1$ and $W_2$. A $morphism$ from $(W_3, Y_3; F)$ to $(W_4, Y_4; G)$ is a module map $\eta$ from $W_3$ to $W_4$ such that

$$G = \overline{\eta} \circ F, \tag{3.5}$$

where $\overline{\eta}$ is the map from $\overline{W}_3$ to $\overline{W}_4$ uniquely extending $\eta$. We define the notion of $P(z)$-tensor product using a universal property as follows:

**Definition 3.1.** A $P(z)$-$tensor$ $product$ $of$ $W_1$ $and$ $W_2$ is a $P(z)$-product

$$(W_1 \boxtimes_{P(z)} W_2, Y_{P(z)}; \boxtimes_{P(z)})$$

such that for any $P(z)$-product $(W_3, Y_3; F)$, there is a unique morphism from

$$(W_1 \boxtimes_{P(z)} W_2, Y_{P(z)}; \boxtimes_{P(z)})$$

to $(W_3, Y_3; F)$. The $V$-module $(W_1 \boxtimes_{P(z)} W_2, Y_{P(z)})$ is called a $P(z)$-$tensor$ $product$ $module$ of $W_1$ and $W_2$.

$Remark$ 3.2. As in the case of tensor products of modules for a Lie algebra, it is clear from the definition that if a $P(z)$-tensor product of $W_1$ and $W_2$ exists, then it is unique up to unique isomorphism.

The existence a $P(z)$-tensor product is not obvious. We shall discuss its existence and constructions under certain conditions, after first relating $P(z)$-intertwining maps to intertwining operators.

Let $\mathcal{Y}$ be an intertwining operator of type $\binom{W_3}{W_1 W_2}$. It follows from the definition of intertwining operator that for any complex number $\zeta$ and any $w_{(1)} \in W_1$,

$$\mathcal{Y}(w_{(1)}, x)\bigg|_{x^n = e^{n\zeta}, \; n \in \mathbb{C}}$$

is a well-defined map from $W_2$ to $\overline{W_3}$. For brevity of notation, we shall write this map as $\mathcal{Y}(w_{(1)}, e^{\zeta})$, but note that $\mathcal{Y}(w_{(1)}, e^{\zeta})$ depends on $\zeta$, not on just $e^{\zeta}$, as the notation might suggest. In this section we shall always choose $\log z$ so that

$$\log z = \log |z| + i \arg z \;\; \text{with} \;\; 0 \le \arg z < 2\pi. \tag{3.6}$$

Arbitrary values of the log function will be denoted

$$l_p(z) = \log z + 2p\pi i \tag{3.7}$$

for $p \in \mathbb{Z}$.

We now describe the precise connection between intertwining operators and $P(z)$-intertwining maps of the same type. Fix an integer $p$. Let $\mathcal{Y}$ be an intertwining operator of type $\binom{W_3}{W_1 W_2}$. We have a linear map $F_{\mathcal{Y},p} : W_1 \otimes W_2 \to \overline{W_3}$ given by

$$F_{\mathcal{Y},p}(w_{(1)} \otimes w_{(2)}) = \mathcal{Y}(w_{(1)}, e^{l_p(z)}) w_{(2)} \tag{3.8}$$

for all $w_{(1)} \in W_1$, $w_{(2)} \in W_2$. Using the Jacobi identity for $\mathcal{Y}$, we see easily that $F_{\mathcal{Y},p}$ is a $P(z)$-intertwining map. Conversely, given a $P(z)$-intertwining map $F$, homogeneous elements $w_{(1)} \in W_1$ and $w_{(2)} \in W_2$ and $n \in \mathbb{C}$, we define $(w_{(1)})_n w_{(2)}$ to be the projection of the image of $w_{(1)} \otimes w_{(2)}$ under $F$ to the homogeneous subspace of $W_3$ of weight

$$\text{wt } w_{(1)} - n - 1 + \text{wt } w_{(2)},$$

multiplied by $e^{(n+1)l_p(z)}$. Using this, we define

$$\mathcal{Y}_{F,p}(w_{(1)}, x) w_{(2)} = \sum_{n \in \mathbb{C}} (w_{(1)})_n w_{(2)} x^{-n-1}, \tag{3.9}$$

and by linearity, we obtain a linear map

$$\begin{aligned} W_1 \otimes W_2 &\to W_3\{x\} \\ w_{(1)} \otimes w_{(2)} &\mapsto \mathcal{Y}_{F,p}(w_{(1)}, x) w_{(2)} \end{aligned}$$

(recall the notation (2.33)).

The proof of the following result is analogous to (and slightly shorter than) the proof of the corresponding result for $Q(z)$-intertwining maps in [HL4]:

**Proposition 3.3.** *For $p \in \mathbb{Z}$, the correspondence $\mathcal{Y} \mapsto F_{\mathcal{Y},p}$ is a linear isomorphism from the vector space $\mathcal{V}_{W_1 W_2}^{W_3}$ of intertwining operators of type $\binom{W_3}{W_1 \, W_2}$ to the vector space $\mathcal{M}[P(z)]_{W_1 W_2}^{W_3}$ of $P(z)$-intertwining maps of type $\binom{W_3}{W_1 W_2}$. Its inverse map is given by $F \mapsto \mathcal{Y}_{F,p}$.*

The following immediate result relates module maps from a tensor product module with intertwining maps and intertwining operators:

**Proposition 3.4.** *Suppose that $W_1 \boxtimes_{P(z)} W_2$ exists. We have a natural isomorphism*

$$\begin{aligned}
\mathrm{Hom}_V(W_1 \boxtimes_{P(z)} W_2, W_3) &\overset{\sim}{\to} \mathcal{M}[P(z)]_{W_1 W_2}^{W_3} \\
\eta &\mapsto \overline{\eta} \circ \boxtimes_{P(z)}
\end{aligned} \tag{3.10}$$

*and for $p \in \mathbb{Z}$, a natural isomorphism*

$$\begin{aligned}
\mathrm{Hom}_V(W_1 \boxtimes_{P(z)} W_2, W_3) &\overset{\sim}{\to} \mathcal{V}_{W_1 W_2}^{W_3} \\
\eta &\mapsto \mathcal{Y}_{\eta,p}
\end{aligned} \tag{3.11}$$

*where $\mathcal{Y}_{\eta,p} = \mathcal{Y}_{F,p}$ with $F = \overline{\eta} \circ \boxtimes_{P(z)}$.*

It is clear from Definition 3.1 that the tensor product operation distributes over direct sums in the following sense:

**Proposition 3.5.** *Let $U_1, \ldots, U_k$, $W_1, \ldots, W_l$ be $V$-modules and suppose that each $U_i \boxtimes_{P(z)} W_j$ exists. Then $(\coprod_i U_i) \boxtimes_{P(z)} (\coprod_j W_j)$ exists and there is a natural isomorphism*

$$\left( \coprod_i U_i \right) \boxtimes_{P(z)} \left( \coprod_j W_j \right) \overset{\sim}{\to} \coprod_{i,j} U_i \boxtimes_{P(z)} W_j. \tag{3.12}$$

Now consider $V$-modules $W_1$, $W_2$ and $W_3$. The natural evaluation map

$$\begin{aligned}
W_1 \otimes W_2 \otimes \mathcal{M}[P(z)]_{W_1 W_2}^{W_3} &\to \overline{W}_3 \\
w_{(1)} \otimes w_{(2)} \otimes F &\mapsto F(w_{(1)} \otimes w_{(2)})
\end{aligned} \tag{3.13}$$

gives a natural map

$$\mathcal{F}_{W_1 W_2}^{W_3} : W_1 \otimes W_2 \to (\mathcal{M}[P(z)]_{W_1 W_2}^{W_3})^* \otimes \overline{W}_3. \tag{3.14}$$

Suppose that $\dim \mathcal{M}[P(z)]_{W_1 W_2}^{W_3} < \infty$. Then $(\mathcal{M}[P(z)]_{W_1 W_2}^{W_3})^* \otimes W_3$ is a $V$-module (with finite-dimensional weight spaces) in the obvious way, and the map $\mathcal{F}_{W_1 W_2}^{W_3}$ is clearly a $P(z)$-intertwining map, where we make the identification

$$(\mathcal{M}[P(z)]_{W_1 W_2}^{W_3})^* \otimes \overline{W}_3 = \overline{(\mathcal{M}[P(z)]_{W_1 W_2}^{W_3})^* \otimes W_3}. \tag{3.15}$$

This gives us a natural $P(z)$-product.

Now we consider a special but important class of vertex operator algebras satisfying certain finiteness and semisimplicity conditions.

**Definition 3.6.** A vertex operator algebra $V$ is *rational* if it satisfies the following conditions:

(1) There are only finitely many irreducible $V$-modules (up to equivalence).
(2) Every $V$-module is completely reducible (and is in particular a *finite* direct sum of irreducible modules).
(3) All the fusion rules (the dimensions of spaces of intertwining operators) for $V$ are finite (for triples of irreducible modules and hence arbitrary modules).

The next result shows that tensor products exist for the category of modules for a rational vertex operator algebra. It is proved by the same argument used to prove the analogous result in [HL4] for the $Q(z)$-tensor product. There is no need to assume that $W_1$ and $W_2$ are irreducible in the formulation or proof, but by Proposition 3.5, the case in which $W_1$ and $W_2$ are irreducible gives all the necessary information, and the tensor product is canonically described using only the spaces of intertwining maps among triples of *irreducible* modules.

**Proposition 3.7.** *Let $V$ be rational and let $W_1$, $W_2$ be $V$-modules. Then*
$$(W_1 \boxtimes_{P(z)} W_2, Y_{P(z)}; \boxtimes_{P(z)})$$
*exists, and in fact*
$$W_1 \boxtimes_{P(z)} W_2 = \coprod_{i=1}^k (\mathcal{M}[P(z)]_{W_1 W_2}^{M_i})^* \otimes M_i, \tag{3.16}$$
*where $\{M_1, \ldots, M_k\}$ is a set of representatives of the equivalence classes of irreducible $V$-modules, and the right-hand side of (3.16) is equipped with the $V$-module and $P(z)$-product structure indicated above. That is,*
$$\boxtimes_{P(z)} = \sum_{i=1}^k \mathcal{F}_{W_1 W_2}^{M_i}. \tag{3.17}$$

*Remark* 3.8. By combining Proposition 3.7 with Proposition 3.3, we can express $W_1 \boxtimes_{P(z)} W_2$ in terms of $\mathcal{V}_{W_1 W_2}^{M_i}$ in place of $\mathcal{M}[P(z)]_{W_1 W_2}^{M_i}$.

The construction in Proposition 3.7 is tautological, and we view the argument as essentially an existence proof. We now give two constructions of a $P(z)$-tensor product (under suitable conditions).

For two $V$-modules $(W_1, Y_1)$ and $(W_2, Y_2)$, we define an action of
$$V \otimes \iota_+ \mathbb{C}[t, t^{-1}, (z^{-1} - t)^{-1}]$$
on $(W_1 \otimes W_2)^*$ (where as in [FLM2] and [HL4], $\iota_+$ denotes the operation of expansion of a rational function of $t$ in the direction of positive powers of $t$), that is, a linear map
$$\tau_{P(z)} : V \otimes \iota_+ \mathbb{C}[t, t^{-1}, (z^{-1} - t)^{-1}] \to \text{End } (W_1 \otimes W_2)^*, \tag{3.18}$$

by

$$
\left( \tau_{P(z)} \left( x_0^{-1} \delta \left( \frac{x_1^{-1} - z}{x_0} \right) Y_t(v, x_1) \right) \lambda \right) (w_{(1)} \otimes w_{(2)})
$$

$$
= z^{-1} \delta \left( \frac{x_1^{-1} - x_0}{z} \right) \lambda(Y_1(e^{x_1 L(1)}(-x_1^{-2})^{L(0)} v, x_0) w_{(1)} \otimes w_{(2)})
$$

$$
+ x_0^{-1} \delta \left( \frac{z - x_1^{-1}}{-x_0} \right) \lambda(w_{(1)} \otimes Y_2^*(v, x_1) w_{(2)}) \tag{3.19}
$$

for $v \in V$, $\lambda \in (W_1 \otimes W_2)^*$, $w_{(1)} \in W_1$, $w_{(2)} \in W_2$, where

$$
Y_t(v, x) = v \otimes x^{-1} \delta \left( \frac{t}{x} \right). \tag{3.20}
$$

The formula (3.19) does indeed give a well-defined map of the type (3.18) (in generating-function form); this definition is motivated by (3.4), in which we first replace $x_1$ by $x_1^{-1}$ and then replace $v$ by $e^{x_1 L(1)}(-x_1^{-2})^{L(0)} v$.

Let $W_3$ be another $V$-module. The space $V \otimes \iota_+ \mathbb{C}[t, t^{-1}, (z^{-1} - t)^{-1}]$ acts on $W_3'$ in the obvious way, where $v \otimes t^n$ ($v \in V$, $n \in \mathbb{Z}$) acts as the component $v_n$ of $Y(v, x)$. The following result, which follows immediately from the definitions (3.4) and (3.19), provides further motivation for the definition of our action on $(W_1 \otimes W_2)^*$:

**Proposition 3.9.** *Under the natural isomorphism*

$$
\mathrm{Hom}(W_3', (W_1 \otimes W_2)^*) \xrightarrow{\sim} \mathrm{Hom}(W_1 \otimes W_2, \overline{W_3}), \tag{3.21}
$$

*the maps in* $\mathrm{Hom}(W_3', (W_1 \otimes W_2)^*)$ *intertwining the two actions of*

$$
V \otimes \iota_+ \mathbb{C}[t, t^{-1}, (z^{-1} - t)^{-1}]
$$

*on $W_3'$ and $(W_1 \otimes W_2)^*$ correspond exactly to the $P(z)$-intertwining maps of type* $\binom{W_3}{W_1 W_2}$.

*Remark* 3.10. Combining the last result with Proposition 3.3, we see that the maps in $\mathrm{Hom}(W_3', (W_1 \otimes W_2)^*)$ intertwining the two actions on $W_3'$ and $(W_1 \otimes W_2)^*$ also correspond exactly to the intertwining operators of type $\binom{W_3}{W_1 W_2}$. In particular, for any intertwining operator $\mathcal{Y}$ of type $\binom{W_3}{W_1 W_2}$ and any integer $p$, the map $F'_{\mathcal{Y},p} : W_3' \to (W_1 \otimes W_2)^*$ defined by

$$
F'_{\mathcal{Y},p}(w'_{(3)})(w_{(1)} \otimes w_{(2)}) = \langle w'_{(3)}, \mathcal{Y}(w_{(1)}, e^{l_p(z)}) w_{(2)} \rangle \tag{3.22}
$$

for $w'_{(3)} \in W_3'$ (recall (3.8); the pairing $\langle \cdot, \cdot \rangle$ on the right-hand side is the obvious one) intertwines the actions of $V \otimes \iota_+ \mathbb{C}[t, t^{-1}, (z^{-1} - t)^{-1}]$ on $W_3'$ and $(W_1 \otimes W_2)^*$.

Write

$$Y'_{P(z)}(v,x) = \tau_{P(z)}(Y_t(v,x)) \tag{3.23}$$

(the specialization of (3.19) to $V \otimes \mathbb{C}[t, t^{-1}]$) and

$$Y'_{P(z)}(\omega,x) = \sum_{n \in \mathbb{Z}} L'_{P(z)}(n) x^{-n-2} \tag{3.24}$$

(recall that $\omega$ is the generator of the Virasoro algebra, for our vertex operator algebra $(V, Y, \mathbf{1}, \omega)$). We call the eigenspaces of the operator $L'_{P(z)}(0)$ the *weight subspaces* or *homogeneous subspaces* of $(W_1 \otimes W_2)^*$, and we have the corresponding notions of *weight vector* (or *homogeneous vector*) and *weight*.

Suppose that $G \in \mathrm{Hom}(W'_3, (W_1 \otimes W_2)^*)$ intertwines the two actions as in Proposition 3.9. Then for $w'_{(3)} \in W'_3$, $G(w'_{(3)})$ satisfies the following nontrivial and subtle condition on $\lambda \in (W_1 \otimes W_2)^*$: The formal Laurent series $Y'_{P(z)}(v, x_0)\lambda$ involves only finitely many negative powers of $x_0$ and

$$\tau_{P(z)}\left(x_0^{-1}\delta\left(\frac{x_1^{-1}-z}{x_0}\right) Y_t(v,x_1)\right)\lambda =$$
$$= x_0^{-1}\delta\left(\frac{x_1^{-1}-z}{x_0}\right) Y'_{P(z)}(v,x_1)\lambda \quad \text{for all } v \in V. \tag{3.25}$$

(Note that the two sides are not *a priori* equal for general $\lambda \in (W_1 \otimes W_2)^*$.) We call this the *compatibility condition* on $\lambda \in (W_1 \otimes W_2)^*$, for the action $\tau_{P(z)}$.

Let $W$ be a subspace of $(W_1 \otimes W_2)^*$. We say that $W$ is *compatible for $\tau_{P(z)}$* if every element of $W$ satisfies the compatibility condition. Also, we say that $W$ is *($\mathbb{C}$-)graded* if it is $\mathbb{C}$-graded by its weight subspaces, and that $W$ is a *$V$-module* (respectively, *generalized module*) if $W$ is graded and is a module (respectively, generalized module) when equipped with this grading and with the action of $Y'_{P(z)}(\cdot, x)$ (recall Definition 2.9). A sum of compatible modules or generalized modules is clearly a generalized module. The weight subspace of a subspace $W$ with weight $n \in \mathbb{C}$ will be denoted $W_{(n)}$.

Given $G$ as above, it is clear that $G(W'_3)$ is a $V$-module since $G$ intertwines the two actions of $V \otimes \mathbb{C}[t, t^{-1}]$. We have in fact established that $G(W'_3)$ is in addition a compatible $V$-module since $G$ intertwines the full actions. Moreover, if $H \in \mathrm{Hom}(W'_3, (W_1 \otimes W_2)^*)$ intertwines the two actions of $V \otimes \mathbb{C}[t, t^{-1}]$, then $H$ intertwines the two actions of $V \otimes \iota_+ \mathbb{C}[t, t^{-1}, (z^{-1}-t)^{-1}]$ if and only if the $V$-module $H(W'_3)$ is compatible.

Define

$$W_1 \boxtimes_{P(z)} W_2 = \sum_{W \in \mathcal{W}_{P(z)}} W = \bigcup_{W \in \mathcal{W}_{P(z)}} W \subset (W_1 \otimes W_2)^*, \tag{3.26}$$

where $\mathcal{W}_{P(z)}$ is the set of all compatible modules for $\tau_{P(z)}$ in $(W_1 \otimes W_2)^*$. Then $W_1 \boxtimes_{P(z)} W_2$ is a compatible generalized module and coincides with the sum (or union) of the images $G(W_3')$ of modules $W_3'$ under the maps $G$ as above. Moreover, for any $V$-module $W_3$ and any map $H : W_3' \to W_1 \boxtimes_{P(z)} W_2$ of generalized modules, $H(W_3')$ is compatible and hence $H$ intertwines the two actions of $V \otimes \iota_+ \mathbb{C}[t, t^{-1}, (z^{-1} - t)^{-1}]$. Thus we have:

**Proposition 3.11.** *The subspace $W_1 \boxtimes_{P(z)} W_2$ of $(W_1 \otimes W_2)^*$ is a generalized module with the following property: Given any $V$-module $W_3$, there is a natural linear isomorphism determined by (3.21) between the space of all $P(z)$-intertwining maps of type $\binom{W_3}{W_1 \, W_2}$ and the space of all maps of generalized modules from $W_3'$ to $W_1 \boxtimes_{P(z)} W_2$.*

For rational vertex operator algebras, we have the following straightforward result:

**Proposition 3.12.** *Let $V$ be a rational vertex operator algebra and $W_1$, $W_2$ two $V$-modules. Then the generalized module $W_1 \boxtimes_{P(z)} W_2$ is a module.*

Now we assume that $W_1 \boxtimes_{P(z)} W_2$ is a module (which occurs if $V$ is rational, by the last proposition). In this case, we define a $V$-module $W_1 \boxtimes_{P(z)} W_2$ by

$$W_1 \boxtimes_{P(z)} W_2 = (W_1 \boxtimes_{P(z)} W_2)' \tag{3.27}$$

(observe the notation $\boxtimes' = \boxtimes$!) and we write the corresponding action as $Y_{P(z)}$. Applying Proposition 3.11 to the special module $W_3 = W_1 \boxtimes_{P(z)} W_2$ and the identity map $W_3' \to W_1 \boxtimes_{P(z)} W_2$, we obtain using (3.21) a canonical $P(z)$-intertwining map of type $\binom{W_1 \boxtimes_{P(z)} W_2}{W_1 \, W_2}$, which we denote

$$\begin{aligned}
\boxtimes_{P(z)} : W_1 \otimes W_2 &\to \overline{W_1 \boxtimes_{P(z)} W_2} \\
w_{(1)} \otimes w_{(2)} &\mapsto w_{(1)} \boxtimes_{P(z)} w_{(2)}.
\end{aligned} \tag{3.28}$$

It is easy to verify:

**Proposition 3.13.** *The $P(z)$-product $(W_1 \boxtimes_{P(z)} W_2, Y_{P(z)}; \boxtimes_{P(z)})$ is a $P(z)$-tensor product of $W_1$ and $W_2$.*

More generally, dropping the assumption that $W_1 \boxtimes_{P(z)} W_2$ is a module, we have:

**Proposition 3.14.** *The $P(z)$-tensor product of $W_1$ and $W_2$ exists (and is given by (3.27)) if and only if $W_1 \boxtimes_{P(z)} W_2$ is a module.*

We observe that any element of $W_1 \boxtimes_{P(z)} W_2$ is an element $\lambda$ of $(W_1 \otimes W_2)^*$ satisfying:

**The compatibility condition (recall (3.25)):**
  **(a)** The *lower truncation condition*: For all $v \in V$, the formal Laurent series $Y'_{P(z)}(v, x)\lambda$ involves only finitely many negative powers of $x$.

**(b)** The following formula holds:

$$\tau_{P(z)}\left(x_0^{-1}\delta\left(\frac{x_1^{-1}-z}{x_0}\right)Y_t(v,x_1)\right)\lambda =$$

$$= x_0^{-1}\delta\left(\frac{x_1^{-1}-z}{x_0}\right)Y'_{P(z)}(v,x_1)\lambda \quad \text{for all } v \in V. \quad (3.29)$$

**The local grading-restriction condition:**

**(a)** The *grading condition*: $\lambda$ is a (finite) sum of weight vectors of $(W_1 \otimes W_2)^*$.

**(b)** Let $W_\lambda$ be the smallest subspace of $(W_1 \otimes W_2)^*$ containing $\lambda$ and stable under the component operators $\tau_{P(z)}(v \otimes t^n)$ of the operators $Y'_{P(z)}(v,x)$ for $v \in V$, $n \in \mathbb{Z}$. Then the weight spaces $(W_\lambda)_{(n)}$, $n \in \mathbb{C}$, of the (graded) space $W_\lambda$ have the properties

$$\dim (W_\lambda)_{(n)} < \infty \quad \text{for } n \in \mathbb{C}, \quad (3.30)$$

$$(W_\lambda)_{(n)} = 0 \quad \text{for } n \text{ whose real part is sufficiently small.} \quad (3.31)$$

We have the following result, which is quite straightforward:

**Proposition 3.15.** *The action $Y'_{P(z)}$ of $V \otimes \mathbb{C}[t,t^{-1}]$ on $(W_1 \otimes W_2)^*$ has the property*

$$Y'_{P(z)}(\mathbf{1},x) = 1, \quad (3.32)$$

*where $1$ on the right-hand side is the identity map of $(W_1 \otimes W_2)^*$, and the $L(-1)$-derivative property*

$$\frac{d}{dx}Y'_{P(z)}(v,x) = Y'_{P(z)}(L(-1)v,x) \quad (3.33)$$

*for $v \in V$.*

The following substantial result can be derived from the corresponding result for $Q(z)$ in place of $P(z)$ in [HL4]–[HL5]:

**Theorem 3.16.** *Let $\lambda$ be an element of $(W_1 \otimes W_2)^*$ satisfying the compatibility condition. Then when acting on $\lambda$, the Jacobi identity for $Y'_{P(z)}$ holds, that is,*

$$x_0^{-1}\delta\left(\frac{x_1-x_2}{x_0}\right)Y'_{P(z)}(u,x_1)Y'_{P(z)}(v,x_2)\lambda$$

$$-x_0^{-1}\delta\left(\frac{x_2-x_1}{-x_0}\right)Y'_{P(z)}(v,x_2)Y'_{P(z)}(u,x_1)\lambda$$

$$= x_2^{-1}\delta\left(\frac{x_1-x_0}{x_2}\right)Y'_{P(z)}(Y(u,x_0)v,x_2)\lambda \quad (3.34)$$

*for $u,v \in V$.*

We also have:

**Proposition 3.17.** *The subspace consisting of the elements of $(W_1 \otimes W_2)^*$ satisfying the compatibility condition is stable under the operators $\tau_{P(z)}(v \otimes t^n)$ for $v \in V$ and $n \in \mathbb{Z}$, and similarly for the subspace consisting of the elements satisfying the local grading-restriction condition.*

We can now formulate the main conclusions. We have another construction of $W_1 \boxtimes_{Q(z)} W_2$:

**Theorem 3.18.** *The subspace of $(W_1 \otimes W_2)^*$ consisting of the elements satisfying the compatibility condition and the local grading-restriction condition, equipped with $Y'_{P(z)}$, is a generalized module and is equal to $W_1 \boxtimes_{P(z)} W_2$.*

The following result follows immediately from Proposition 3.14, the theorem above and the definition of $W_1 \boxtimes_{Q(z)} W_2$:

**Corollary 3.19.** *The $P(z)$-tensor product of $W_1$ and $W_2$ exists if and only if the subspace of $(W_1 \otimes W_2)^*$ consisting of the elements satisfying the compatibility condition and the local grading-restriction condition, equipped with $Y'_{P(z)}$, is a module. In this case, this module coincides with the module $W_1 \boxtimes_{P(z)} W_2$, and the contragredient module of this module, equipped with the $P(z)$-intertwining map $\boxtimes_{P(z)}$, is a $P(z)$-tensor product of $W_1$ and $W_2$, equal to the structure $(W_1 \boxtimes_{P(z)} W_2, Y_{P(z)}; \boxtimes_{P(z)})$ constructed above.*

From this result and Propositions 3.12 and 3.13, we have:

**Corollary 3.20.** *Let $V$ be a rational vertex operator algebra and $W_1$, $W_2$ two $V$-modules. Then the $P(z)$-tensor product $(W_1 \boxtimes_{P(z)} W_2, Y_{P(z)}; \boxtimes_{P(z)})$ may be constructed as described in Corollary 3.19.*

## 4. VERTEX TENSOR CATEGORIES

In the preceding section, we have introduced the notion of $P(z)$-tensor product, exhibited some straightforward consequences and presented two constructions, the second of which is the important one. This theory of tensor products is in fact much richer than what we have seen in Section 3. In this section, instead of giving a complete description of the theory (which is the task of the series of papers mentioned in the introduction), we summarize the properties of tensor products of modules for a rational vertex operator algebra under the condition that products of intertwining operators converge in a suitable sense: We define the notion of vertex tensor category, and we announce that the category of modules for such an algebra gives the structure of a vertex tensor category and also that any vertex tensor category gives the structure of a braided tensor category.

To define the notion of vertex tensor category, we need the moduli space $K$ and the determinant line bundle over $K$ mentioned at the beginning of the preceding section. We also use the language of (partial) operads. For details, see [M], [H1]–[H5] and [HL2]–[HL3]. As in these papers, $\tilde{K}^1$ is the determinant line bundle over the moduli space $K$, and for any complex number $c$, $\tilde{K}^c$ is the determinant line bundle $\tilde{K}^1$ raised to the power $c$. Then $K$, $\tilde{K}^1$ and $\tilde{K}^c$ for any $c \in \mathbb{C}$ are all $\mathbb{C}^\times$-rescalable associative analytic partial operads (see [HL2], [HL3], [H5]). We use the notations $K(n)$, $\tilde{K}^1(n)$ and $\tilde{K}^c(n)$ for $n \in \mathbb{N}$ to denote the $n$th components of these operadic structures (corresponding to spheres with $n$ positively oriented punctures and one negatively oriented puncture). These components are path-connected. The substitution (composition) map for the partial operad $\tilde{K}^c$ is denoted by $\gamma$; this map describes the sewing operation. The identity element (of $\tilde{K}^c(1)$) is denoted by $\tilde{I}$. Below we always use the convention that when we write an expression involving the substitution map of $\tilde{K}^c$, we have assumed that the expression exists. We also need the flat section $\psi : K \to \tilde{K}^c$, given in [H2] and [H5], used to obtain the vertex operators and the central charge for the vertex operator algebra corresponding to a vertex associative algebra. We denote by $J$ the element of $K(0)$ containing the sphere $\mathbb{C} \cup \{0\}$ with the negatively oriented puncture $\infty$ and the standard local coordinate vanishing at $\infty$. Also recall the notation $P(z)$ from the preceding section.

**Definition 4.1.** A *vertex tensor (monoidal) category of central charge* $c \in \mathbb{C}$ consists of a category $\mathcal{V}$ together with the following data (see below for the axioms):

(1) For every $\tilde{Q} \in \tilde{K}^c(2)$, we have a bifunctor

$$\boxtimes_{\tilde{Q}} : \mathcal{V} \times \mathcal{V} \to \mathcal{V}, \tag{4.1}$$

called the $\tilde{Q}$-*tensor product bifunctor*.

(2) For any $\tilde{Q} \in \tilde{K}^c(1)$, we have a functor

$$e_{\tilde{Q}} : \mathcal{V} \to \mathcal{V}, \tag{4.2}$$

called the $\tilde{Q}$-*quasi-identity functor*, such that

$$e_{\tilde{I}} = 1, \tag{4.3}$$

where 1 is the identity functor from $\mathcal{V}$ to $\mathcal{V}$.

(3) We have an object $V$ in $\mathcal{V}$, called the *unit object*.

(4) For any $\tilde{Q}_1$, $\tilde{Q}_2$, $\tilde{Q}_3$, $\tilde{Q}_4 \in \tilde{K}^c(2)$ satisfying

$$\gamma(\tilde{Q}_1; \tilde{Q}_2, \tilde{I}) = \gamma(\tilde{Q}_3; \tilde{I}, \tilde{Q}_4), \tag{4.4}$$

we have a natural isomorphism

$$\mathcal{A}^{\tilde{Q}_3, \tilde{Q}_4}_{\tilde{Q}_1, \tilde{Q}_2} : \boxtimes_{\tilde{Q}_1} \circ (\boxtimes_{\tilde{Q}_2} \times 1) \to \boxtimes_{\tilde{Q}_3} \circ (1 \times \boxtimes_{\tilde{Q}_4}) \tag{4.5}$$

(where as above 1 is the identity functor from $\mathcal{V}$ to $\mathcal{V}$), called the *associativity isomorphism*, so that in particular, for any objects $W_1$, $W_2$, $W_3$ in $\mathcal{V}$ we have an associativity isomorphism

$$(W_1 \boxtimes_{\tilde{Q}_2} W_2) \boxtimes_{\tilde{Q}_1} W_3 \to W_1 \boxtimes_{\tilde{Q}_3} (W_2 \boxtimes_{\tilde{Q}_4} W_3). \tag{4.6}$$

(5) For any $\tilde{Q} \in \tilde{K}^c(2)$, we have a natural isomorphism

$$\mathcal{C}_{\tilde{Q}} : \boxtimes_{\tilde{Q}} \to \boxtimes_{\sigma_{12}(\tilde{Q})} \circ \sigma_{12}, \tag{4.7}$$

called the *commutativity isomorphism*, where $\sigma_{12}$ denotes the nontrivial element of $S_2$ and also the functor from $\mathcal{V} \times \mathcal{V}$ to itself given by the permutation $\sigma_{12}$, so that in particular, for any objects $W_1, W_2$ in $\mathcal{V}$ we have a commutativity isomorphism

$$W_1 \boxtimes_{\tilde{Q}} W_2 \to W_2 \boxtimes_{\sigma_{12}(\tilde{Q})} W_1. \tag{4.8}$$

(6) For any $\tilde{Q}_1, \tilde{Q}_2 \in \tilde{K}^c(2)$ and any homotopy class $[\Gamma]$ of paths $\Gamma$ from $\tilde{Q}_1$ to $\tilde{Q}_2$, we have a natural isomorphism

$$\mathcal{T}_{[\Gamma]} : \boxtimes_{\tilde{Q}_1} \to \boxtimes_{\tilde{Q}_2}, \tag{4.9}$$

called the *parallel transport isomorphism*, so that in particular, for any objects $W_1, W_2$ in $\mathcal{V}$ we have a parallel transport isomorphism

$$W_1 \boxtimes_{\tilde{Q}_1} W_2 \to W_1 \boxtimes_{\tilde{Q}_2} W_2. \tag{4.10}$$

(7) Let $\tilde{Q}_1, \tilde{Q}_2, \tilde{Q}_3 \in \tilde{K}^c(1)$ and $\tilde{R}_1, \tilde{R}_2 \in \tilde{K}^c(2)$ and suppose that either

$$\gamma(\tilde{Q}_1; \tilde{R}_1) = \tilde{R}_2, \tag{4.11}$$

$$\gamma(\tilde{R}_1; \tilde{Q}_1, \tilde{I}) = \tilde{R}_2, \tag{4.12}$$

$$\gamma(\tilde{R}_1; \tilde{I}, \tilde{Q}_1) = \tilde{R}_2 \tag{4.13}$$

or

$$\gamma(\tilde{Q}_1; \tilde{Q}_2) = \tilde{Q}_3. \tag{4.14}$$

Then we have a natural isomorphism

$$\mathcal{S}[0]^{\tilde{R}_2}_{\tilde{Q}_1, \tilde{R}_1} : e_{\tilde{Q}_1} \circ \boxtimes_{\tilde{R}_1} \to \boxtimes_{\tilde{R}_2}, \tag{4.15}$$

$$\mathcal{S}[1]^{\tilde{R}_2}_{\tilde{R}_1, \tilde{Q}_1} : \boxtimes_{\tilde{R}_1} \circ (e_{\tilde{Q}_1} \times 1) \to \boxtimes_{\tilde{R}_2}, \tag{4.16}$$

$$\mathcal{S}[2]^{\tilde{R}_2}_{\tilde{R}_1, \tilde{Q}_1} : \boxtimes_{\tilde{R}_1} \circ (1 \times e_{\tilde{Q}_1}) \to \boxtimes_{\tilde{R}_2} \tag{4.17}$$

or

$$\mathcal{S}^{\tilde{Q}_3}_{\tilde{Q}_1, \tilde{Q}_2} : e_{\tilde{Q}_1} \circ e_{\tilde{Q}_2} \to e_{\tilde{Q}_3}, \tag{4.18}$$

respectively, called a *substitution isomorphism*, so that in particular, for any objects $W_1$, $W_2$ in $\mathcal{V}$ we have the corresponding substitution isomorphism

$$e_{\tilde{Q}_1}(W_1 \boxtimes_{\tilde{R}_1} W_2) \to W_1 \boxtimes_{\tilde{R}_2} W_2, \tag{4.19}$$

$$e_{\tilde{Q}_1}(W_1) \boxtimes_{\tilde{R}_1} W_2 \to W_1 \boxtimes_{\tilde{R}_2} W_2, \tag{4.20}$$

$$W_1 \boxtimes_{\tilde{R}_1} e_{\tilde{Q}_1}(W_2) \to W_1 \boxtimes_{\tilde{R}_2} W_2 \tag{4.21}$$

or

$$e_{\tilde{Q}_1}(e_{\tilde{Q}_2}(W_1)) \to e_{\tilde{Q}_3}(W_1), \tag{4.22}$$

respectively.

(8) Let $\tilde{Q}_1 \in \check{K}^c(1)$, $\tilde{Q}_2 \in \check{K}^c(2)$ and suppose that either

$$\gamma(\tilde{Q}_2; \psi(J), \tilde{I}) = \tilde{Q}_1 \tag{4.23}$$

or

$$\gamma(\tilde{Q}_2; \tilde{I}, \psi(J)) = \tilde{Q}_1. \tag{4.24}$$

Then we have a natural isomorphism (the *left unit isomorphism*)

$$\mathcal{L}^{\tilde{Q}_1}_{\tilde{Q}_2} : V \boxtimes_{\tilde{Q}_2} \cdot \to e_{\tilde{Q}_1}, \tag{4.25}$$

or (the *right unit isomorphism*)

$$\mathcal{R}^{\tilde{Q}_1}_{\tilde{Q}_2} : \cdot \boxtimes_{\tilde{Q}_2} V \to e_{\tilde{Q}_1}, \tag{4.26}$$

respectively, where $V \boxtimes_{\tilde{Q}_2} \cdot$ and $\cdot \boxtimes_{\tilde{Q}_2} V$ are the functors from $\mathcal{V}$ to itself which take an object $W$ to $V \boxtimes_{\tilde{Q}_2} W$ and $W \boxtimes_{\tilde{Q}_2} V$, respectively, and a morphism $f \in \mathrm{Hom}(W_1, W_2)$ to $1_V \boxtimes_{\tilde{Q}_2} f$ and $f \boxtimes_{\tilde{Q}_2} 1_V$, respectively, so that in particular, for any object $W$ in $\mathcal{V}$ we have two unit isomorphisms

$$V \boxtimes_{\tilde{Q}_2} W \to e_{\tilde{Q}_1}(W) \tag{4.27}$$

or

$$W \boxtimes_{\tilde{Q}_2} V \to e_{\tilde{Q}_1}(W), \tag{4.28}$$

respectively.

These data are required to satisfy the following axioms:

(1) The symmetry condition

$$\mathcal{C}_{\sigma_{12}(\tilde{Q})} \circ \mathcal{C}_{\tilde{Q}} = \mathcal{T}_{[\Gamma_{\tilde{Q}}]} \tag{4.29}$$

holds, where $\tilde{Q} \in \check{K}^c(2)$ and $[\Gamma_{\tilde{Q}}]$ is the homotopy class of loops based at $\tilde{Q}$ homotopic to the loop $\Gamma_{\tilde{Q}}$ given by $\{\psi(P(e^{2\pi it}))|0 \le t \le 1\}$.

(2) For any $n \in \mathbb{N}$, consider the directed graph $\mathcal{G}_n$ whose vertices are the decompositions of the elements of $\tilde{K}^c(n)$ into $\psi(J)$ and elements of $\tilde{K}^c(1)$ and $\tilde{K}^c(2)$ using the substitution map $\gamma$ (the sewing operation), and whose arrows are induced from the equality (4.4), from permutations of the orderings of the two positively oriented punctures of elements of $\tilde{K}^c(2)$, from homotopy classes of paths connecting elements of $\tilde{K}^c(2)$, from the equalities (4.11)–(4.14), and finally, from the equalities (4.23)–(4.24). We require the following: Let $D_1$ and $D_2$ be vertices of $\mathcal{G}_n$ and let $c_1$ and $c_2$ be chains of arrows from $D_1$ to $D_2$, giving us a diagram in $\mathcal{G}_n$. We have a corresponding diagram whose vertices are obtained by replacing the vertices of the original diagram by multifunctors constructed using the tensor product bifunctors, the quasi-identity functors, the unit object and the decompositions of elements of $\tilde{K}^c(n)$ given by the original vertices, and whose arrows are the natural isomorphisms induced from the associativity, commutativity, parallel transport, substitution and unit isomorphisms corresponding to the arrows in the original diagram. Then this diagram of multifunctors commutes, up to the natural automorphism of the multifunctor corresponding to $D_2$ associated with the homotopy class of loops in $\tilde{K}^c(n)$ obtained in the obvious way from the chains $c_1$ and $c_2$.

The following theorem is the main result of our theory:

**Theorem 4.2.** *Let $V$ be a rational vertex operator algebra of central charge $c \in \mathbb{C}$ and assume that the product of any $n$ formally composable intertwining operators $\mathcal{Y}_1(w_{(1)}, x_1), \ldots, \mathcal{Y}_n(w_{(n)}, x_n)$ is absolutely convergent when $x_1^m, \ldots, x_n^m$, $m \in \mathbb{C}$, are replaced by the complex numbers $e^{m \log z_1}, \ldots, e^{m \log z_n}$ (recall (3.6)) for any $z_1, \ldots, z_n \in \mathbb{C}$ satisfying $|z_1| > \cdots > |z_n| > 0$. Then the category of $V$-modules has a natural structure of vertex tensor category of central charge $c$.*

In particular, for the familiar vertex operator algebras, we have:

**Corollary 4.3.** *For the vertex operator algebra $V$ associated with a WZNW model of positive integral level $k$, a minimal model of central charge $1 - 6\frac{(p-q)^2}{pq}$ (where $p, q$ are relatively prime positive integers) or the moonshine module, the category of $V$-modules has a natural structure of vertex tensor category of central charge $\frac{kd}{k+h}$ (where $d$ is the dimension of the finite-dimensional simple Lie algebra associated with the WZNW model and $h$ is the dual Coxeter number), $1 - 6\frac{(p-q)^2}{pq}$ or $24$, repectively.*

Even in the case of the moonshine module, this vertex tensor category structure is nontrivial.

For tensor (monoidal) categories, there is a coherence theorem [ML]. For vertex tensor categories, we also have a coherence theorem, which reduces the verification of the axiom (2) in the definition of vertex tensor category to the commutativity of certain simple diagrams.

We now describe the construction of a braided tensor (monoidal) category (see [JS1]) from a vertex tensor category. Let $\mathcal{V}$ be a vertex tensor category. We first define the tensor product bifunctor to be

$$\boxtimes = \boxtimes_{\psi(P(1))}, \tag{4.30}$$

so that for two objects $W_1$ and $W_2$ in the underlying category of $\mathcal{V}$, we have a tensor product

$$W_1 \boxtimes W_2 = W_1 \boxtimes_{\psi(P(1))} W_2. \tag{4.31}$$

The associativity isomorphism $a$ is defined by

$$a = \mathcal{T}_{[\Gamma_1]} \circ \mathcal{A}^{\psi(P(2)),\psi(P(1))}_{\psi(P(1)),\psi(P(1))} : \boxtimes \circ (\boxtimes \times 1) \to \boxtimes \circ (1 \times \boxtimes), \tag{4.32}$$

where $\Gamma_1$ is a fixed path from $\psi(P(2))$ to $\psi(P(1))$ in $\tilde{K}^c(2)$ and where we view $\mathcal{T}_{[\Gamma_1]}$ as a natural transformation from $\boxtimes_{\psi(P(2))} \circ (1 \times \boxtimes_{\psi(P(1))})$ to $\boxtimes_{\psi(P(1))} \circ (1 \times \boxtimes_{\psi(P(1))})$. The unit object is $V$. The unit isomorphisms are defined by

$$l = \mathcal{L}^{\tilde{I}}_{\psi(P(1))} : V \boxtimes \cdot \to 1, \tag{4.33}$$

$$r = \mathcal{R}^{\tilde{I}}_{\sigma_{12}(\psi(P(1)))} \circ \mathcal{T}_{[\Gamma_2]} : \cdot \boxtimes V \to 1, \tag{4.34}$$

where $\Gamma_2$ is a fixed path from $\psi(P(1))$ to $\sigma_{12}(\psi(P(1)))$. The braiding isomorphism is defined by

$$c = \mathcal{T}_{[\Gamma_3]} \circ \mathcal{C}_{\psi(P(1))} : \boxtimes \to \boxtimes \circ \sigma_{12}, \tag{4.35}$$

where $\Gamma_3$ is a fixed path from $\sigma_{12}(\psi(P(1)))$ to $\psi(P(1))$.

**Theorem 4.4.** *The underlying category of $\mathcal{V}$, together with the tensor product $\boxtimes$ and the other data defined above, forms a braided tensor category. Different choices of $[\Gamma_1], [\Gamma_2]$ and $[\Gamma_3]$ give isomorphic braided tensor categories.*

In particular, we have:

**Corollary 4.5.** *For the vertex operator algebra associated with a WZNW model, a minimal model or the moonshine module, the category of modules has a natural structure of braided tensor category.*

These braided tensor categories yield the usual fusion rules and the usual fusing and braiding structures studied in conformal field theory (see [BPZ], [Ve], [MS], [BNS], [Fi], etc.). For the moonshine module, this braided tensor category is trivial, although, as we have mentioned above, the vertex tensor category structure which produces it is nontrivial.

## References

[BPZ]   A. A. Belavin, A. M. Polyakov and A. B. Zamolodchikov, Infinite conformal symmetries in two-dimensional quantum field theory, *Nucl. Phys.* **B241** (1984), 333–380.

[B1]    R. E. Borcherds, Vertex algebras, Kac-Moody algebras, and the Monster, *Proc. Natl. Acad. Sci. USA* **83** (1986), 3068–3071.

[B2]    R. E. Borcherds, Monstrous moonshine and monstrous Lie superalgebras, *Invent. Math.* **109** (1992), 405–444.

[BNS]   R. Brustein, Y. Ne'eman and S. Sternberg, Duality, crossing and Mac Lane's coherence.

[CN]    J. H. Conway and S. P. Norton, Monstrous moonshine, *Bull. London Math. Soc.* **11** (1979), 308–339.

[Do]    C. Dong, Representations of the moonshine module vertex operator algebra, in: *Proc. 1992 Joint Summer Research Conference on Conformal Field Theory, Topological Field Theory and Quantum Groups, Mt. Holyoke, 1992*, Contemporary Math., Amer. Math. Soc., Providence, to appear.

[DL]    C. Dong and J. Lepowsky, *Generalized Vertex Algebras and Relative Vertex Operators*, Progress in Math., Vol. 112, Birkhäuser, Boston, 1993.

[Dr1]   V. Drinfeld, On quasi-cocommutative Hopf algebras, *Algebra and Analysis* **1** (1989), 30–46.

[Dr2]   V. Drinfeld, On quasitriangular quasi-Hopf algebras and a certain group closely related to Gal($\bar{\mathbb{Q}}/\mathbb{Q}$), *Algebra and Analysis* **2** (1990), 149–181.

[Fa]    G. Faltings, A proof for the Verlinde formula, to appear.

[Fi]    M. Finkelberg, Fusion categories, Ph.D. thesis, Harvard University, 1993.

[FHL]   I. B. Frenkel, Y.-Z. Huang and J. Lepowsky, On axiomatic approaches to vertex operator algebras and modules, preprint, 1989; *Memoirs Amer. Math. Soc.* **104**, Number 494, 1993.

[FLM1]  I. B. Frenkel, J. Lepowsky and A. Meurman, A natural representation of the Fischer-Griess Monster with the modular function $J$ as character, *Proc. Natl. Acad. Sci. USA* **81** (1984), 3256–3260.

[FLM2]  I. B. Frenkel, J. Lepowsky and A. Meurman, *Vertex Operator Algebras and the Monster*, Pure and Appl. Math., Vol. 134, Academic Press, Boston, 1988.

[FZ]    I. B. Frenkel and Y. Zhu, Vertex operator algebras associated to representations of affine and Virasoro algebras, *Duke Math. J.* **66** (1992), 123–168.

[FS]    D. Friedan and S. Shenker, The analytic geometry of two-dimensional conformal field theory, *Nucl. Phys.* **B281** (1987), 509–545.

[Go]    D. Gorenstein, *Finite Simple Groups. An Introduction to Their Classification*, Plenum Press, New York, 1982.

[Gr]    R. L. Griess, Jr., The Friendly Giant, *Invent. Math.* **69** (1982), 1–102.

[H1]    Y.-Z. Huang, *On the geometric interpretation of vertex operator algebras*, Ph.D. thesis, Rutgers University, 1990; Operads and the geometric interpretation of vertex operator algebras, I, to appear.

[H2]    Y.-Z. Huang, Geometric interpretation of vertex operator algebras, *Proc. Natl. Acad. Sci. USA* **88** (1991), 9964–9968.

[H3]    Y.-Z. Huang, Applications of the geometric interpretation of vertex operator algebras, in: *Proc. 20th International Conference on Differential Geometric Methods in Theoretical Physics, New York, 1991*, ed. S. Catto and A. Rocha, World Scientific, Singapore, 1992, Vol. 1, 333–343.

[H4]    Y.-Z. Huang, Vertex operator algebras and conformal field theory, *Intl. J. Mod. Phys.* **A7** (1992), 2109–2151.

[H5]    Y.-Z. Huang, Operads and the geometric interpretation of vertex operator algebras, II, in

preparation.

[HL1]   Y.-Z. Huang and J. Lepowsky, Toward a theory of tensor products for representations of a vertex operator algebra, in: *Proc. 20th International Conference on Differential Geometric Methods in Theoretical Physics, New York, 1991*, ed. S. Catto and A. Rocha, World Scientific, Singapore, 1992, Vol. 1, 344–354.

[HL2]   Y.-Z. Huang and J. Lepowsky, Vertex operator algebras and operads, in: *The Gelfand Mathematical Seminars, 1990–1992*, ed. L. Corwin, I. Gelfand and J. Lepowsky, Birkhäuser, Boston, 1993, 145–161.

[HL3]   Y.-Z. Huang and J. Lepowsky, Operadic formulation of the notion of vertex operator algebra, in: *Proc. 1992 Joint Summer Research Conference on Conformal Field Theory, Topological Field Theory and Quantum Groups, Mt. Holyoke, 1992*, Contemporary Math., Amer. Math. Soc., Providence, to appear.

[HL4]   Y.-Z. Huang and J. Lepowsky, A theory of tensor products for module categories for a vertex operator algebra, I, to appear.

[HL5]   Y.-Z. Huang and J. Lepowsky, A theory of tensor products for module categories for a vertex operator algebra, II, to appear.

[J]     V. F. R. Jones, Hecke algebra representations of braid groups and link polynomials, *Ann. Math.* **126** (1987), 335–388.

[JS1]   A. Joyal and R. Street, Braided monoidal categories, *Macquarie Mathematics Reports*, Macquarie University, Australia, 1986.

[JS2]   A. Joyal and R. Street, An introduction to Tannaka duality and quantum groups, in: *Category theory, Como, 1990*, ed. A. Carboni, M.C. Pedicchio and G. Rosolini, Lecture Notes in Math. **1488**, Springer-Verlag, Berlin, 1991, 413–492.

[KL1]   D. Kazhdan and G. Lusztig, Affine Lie algebras and quantum groups, *International Mathematics Research Notices* (in *Duke Math. J.*) **2** (1991), 21–29.

[KL2]   D. Kazhdan and G. Lusztig, Tensor structures arising from affine Lie algebras, I, *J. Amer. Math. Soc.* **6** (1993), 905–947.

[KL3]   D. Kazhdan and G. Lusztig, Tensor structures arising from affine Lie algebras, II, *J. Amer. Math. Soc.* **6** (1993), 949–1011.

[KL4]   D. Kazhdan and G. Lusztig, Tensor structures arising from affine Lie algebras, III, to appear.

[KL5]   D. Kazhdan and G. Lusztig, Tensor structures arising from affine Lie algebras, IV, to appear.

[KZ]    V. G. Knizhnik and A. B. Zamolodchikov, Current algebra and Wess-Zumino models in two dimensions, *Nucl. Phys.* **B247** (1984), 83–103.

[K1]    T. Kohno, Linear representations of braid groups and classical Yang-Baxter equations, in: *Braids, Santa Cruz, 1986*, *Contemporary Math.* **78** (1988), 339–363.

[K2]    T. Kohno, Monodromy representations of braid groups and Yang-Baxter equations, *Ann. Inst. Fourier* **37** (1987), 139–160.

[L1]    J. Lepowsky, Perspectives on vertex operators and the Monster, in: Proc. 1987 Symposium on the Mathematical Heritage of Herman Weyl, Duke Univ., *Proc. Symp. Pure Math., Amer. Math. Soc.* **48** (1988), 181–197.

[L2]    J. Lepowsky, Remarks on vertex operator algebras and moonshine, in: *Proc. 20th International Conference on Differential Geometric Methods in Theoretical Physics, New York, 1991*, ed. S. Catto and A. Rocha, World Scientific, Singapore, 1992, Vol. 1, 362–370.

[LW]    J. Lepowsky and R.L. Wilson, A new family of algebras underlying the Rogers-Ramanujan identities and generalizations, *Proc. Natl. Acad. Sci. USA* **78** (1981), 7254–7258.

[ML]    S. Mac Lane, *Categories for the Working Mathematician*, Graduate Texts in Math., Vol. 5, Springer-Verlag, New York, 1971.

[MN]    G. Mason, Finite groups and modular functions (with an appendix by S. P. Norton), in:

Representations of Finite Groups, Proc. 1986 Summer Research Institute, Arcata, ed. P. Fong, *Proc. Symp. Pure Math., Amer. Math. Soc.* **47** (1987), 181–210.

[M]     J. P. May, *The geometry of iterated loop spaces*, Lecture Notes in Math. **271**, Springer-Verlag, 1972.

[MS]    G. Moore and N. Seiberg, Classical and quantum conformal field theory, *Comm. Math. Phys.* **123** (1989), 177–254.

[RT]    N. Reshetikhin and V. G. Turaev, Invariants of 3-manifolds via link polynomials and quantum groups, *Invent. Math.* **103** (1991), 547–597.

[SV]    V. V. Schechtman and A. N. Varchenko, Arrangements of hyperplanes and Lie algebra homology, *Invent. Math.* **106** (1991), 139–194.

[S1]    J. D. Stasheff, Homotopy associativity of *H*-spaces, I, *Trans. Amer. Math. Soc.* **108** (1963), 275–292.

[S2]    J. D. Stasheff, Homotopy associativity of *H*-spaces, II, *Trans. Amer. Math. Soc.* **108** (1963), 293–312.

[TK]    A. Tsuchiya and Y. Kanie, Vertex operators in conformal field theory on $\mathbb{P}^1$ and monodromy representations of braid group, in: *Conformal Field Theory and Solvable Lattice Models, Advanced Studies in Pure Math.*, Vol. 16, Kinokuniya Company Ltd., Tokyo, 1988, 297–372.

[TUY]   A. Tsuchiya, K. Ueno and Y. Yamada, Conformal field theory on universal family of stable curves with gauge symmetries, in: *Advanced Studies in Pure Math.*, Vol. 19, Kinokuniya Company Ltd., Tokyo, 1989, 459–565.

[Va]    A. N. Varchenko, Hypergeometric functions and representation theory of Lie algebras and quantum groups, to appear.

[Ve]    E. Verlinde, Fusion rules and modular transformations in 2D conformal field theory, *Nucl. Phys.* **B300** (1988), 360–376.

[Wa]    W. Wang, Rationality of Virasoro vertex operator algebras, *International Mathematics Research Notices* (in *Duke Math. J.*) **7** (1993), 197–211.

[Wi1]   E. Witten, Non-abelian bosonization in two dimensions, *Comm. Math. Phys.* **92** (1984), 455–472.

[Wi2]   E. Witten, Quantum field theory and the Jones polynomial, *Comm. Math. Phys.* **121** (1989), 351–399.

DEPARTMENT OF MATHEMATICS, UNIVERSITY OF PENNSYLVANIA, PHILADELPHIA, PA 19104
*E-mail address*: yzhuang@math.upenn.edu

DEPARTMENT OF MATHEMATICS, RUTGERS UNIVERSITY, NEW BRUNSWICK, NJ 08903
*E-mail address*: lepowsky@math.rutgers.edu